\documentclass[aps,prx,nofootinbib,twocolumn,amsmath,amssymb,superscriptaddress,notitlepage,groupedaddress]{revtex4-1}

\usepackage{latexsym}
\usepackage{graphicx}
\usepackage{times,psfrag,subfigure}
\usepackage{enumerate}
\usepackage{amsmath}
\usepackage{dsfont}
\usepackage{dcolumn}
\usepackage{bm,bbm}
\usepackage{color}
\usepackage{latexsym,amsmath,amssymb,bm,euscript}
\usepackage{dsfont}
\usepackage{textcomp}
\usepackage{tabularx}
\usepackage{setspace}
\usepackage{ctable}
\usepackage{sidecap}
\usepackage{placeins}
\usepackage{threeparttable}
\usepackage{multirow}

\let \oldbm \bm
\renewcommand{\vec}[1]{\oldbm{#1}}

\def\bk{{\vec k}}
\def\bD{{\vec D}}
\def\ba{{\vec a}}

\def\bq{{\vec q}}

\def\bR{{\vec R}}
\def\bG{{\vec G}}
\def\bd{{\vec d}}

\def\bp{{\vec p}}

\def\bd{{\vec d}}
\def\br{{\vec r}}

\def\bsigma{{\boldsymbol \sigma}}
\def\bnabla{{\boldsymbol \nabla}}

\def\D{\mathcal{D}}

\def\H{\mathcal{H}}

\def\M{\mathcal{M}}

\def\diag{{\rm diag}}

\newcommand{\beq}{\begin{equation}}
\newcommand{\eeq}{\end{equation}}
\newcommand{\beqarray}{\begin{eqnarray}}
\newcommand{\eeqarray}{\end{eqnarray}}

\allowdisplaybreaks

\graphicspath{{figures/}}

\begin{document}

\title{Magic Angle Hierarchy in Twisted Graphene Multilayers}

\author{Eslam Khalaf}
\email{eslam\_khalaf@fas.harvard.edu}
\affiliation{Department of Physics, Harvard University, Cambridge, MA 02138}

\author{Alex J. Kruchkov}
\affiliation{Department of Physics, Harvard University, Cambridge, MA 02138}

\author{Grigory Tarnopolsky}
\affiliation{Department of Physics, Harvard University, Cambridge, MA 02138}

\author{Ashvin Vishwanath}
\affiliation{Department of Physics, Harvard University, Cambridge, MA 02138}

\date{\today}

\begin{abstract}

When two monolayers of graphene are stacked with a small relative twist angle, the resulting band structure exhibits a remarkably flat pair of bands at a sequence of 'magic angles' where correlation effects can induce a host of exotic phases. Here, we study a class of related models of $n$-layered graphene with alternating relative twist angle $\pm \theta$ which exhibit magic angle flat bands coexisting with several Dirac dispersing bands at the Moir\'e K point. Remarkably, we find that the Hamiltonian for the multilayer system can be mapped exactly to a set of decoupled bilayers at {\it different} angles, revealing a remarkable hierarchy mathematically relating all these magic angles to the TBG case. For the trilayer case ($n = 3$), we show that the sequence of magic angle is obtained by multiplying the bilayer magic angles by $\sqrt{2}$, whereas the quadrilayer case ($n = 4$) has two sequences of magic angles obtained by multiplying the bilayer magic angles by the golden ratio $\varphi = (\sqrt{5} + 1)/2 \approx 1.62$ and its inverse. We also show that for larger $n$, we can tune the angle to obtain several narrow (almost flat) bands simultaneously and that for $n \rightarrow \infty$, there is a continuum of magic angles for $\theta \lesssim 2^o$. Furthermore, we show that tuning several perfectly flat bands for a small number of layers is possible if the coupling between different layers is different. The setup proposed here can be readily achieved by repeatedly applying the "tear and stack" method without the need of any extra tuning of the twist angle and has the advantage that the first magic angle is always larger than the bilayer case.

\end{abstract}

\maketitle

\section{Introduction}
Recently, it was shown that two graphene layers twisted to a special ("magic") angle exhibit a very interesting range of correlated phenomena including Mott insulating and superconducting phases \cite{Cao2018a,Cao2018b,Yankowitz2018}. This remarkable discovery has stimulated further extensive research into magic-angle superconductivity and correlated electron states in van der Waals heterostructures \cite{Po2018,Thomson2018,Zou2018,Guinea2018a,Carr2018a,Su2018,Gonzalez2018,Wu2018b,
Efimkin2018,Yuan2018,Xu2018,Ochi2018,Wu2018a,Zhang2018,Wu2018,Kang2018,Pizarro2018,
Koshino2018,Kennes2018,Isobe2018,Rademaker2018,Qiao2018,Chung2018, Fidrysiak2018, Peltonen2018,Tarnopolsky2018,Bernevig2018,Balents2018, Po2018a, TDBGexp2019, IOP_TDBG, PabloTDBG, Lee2019theory} and inspired a vast theoretical and experimental search to extend the family of systems which exhibit similar behavior  \cite{Amorim2018,Zhang2018,Zuo2018,Chen2019,TDBGexp2019, IOP_TDBG, PabloTDBG, Lee2019theory}. Finding such systems achieves several goals. First, they expand the family of Moir\'e systems where correlated physics can be studied in a setting which have several advantages over traditional strongly-correlated systems (easier to fabricate, richer possibilities for  tuning the band structures, etc). Second, finding systems which share similarities with TBG,  but differ in some details --  such as symmetries, bands topology and interaction strength, -- can help provide a deeper understanding of the correlated physics in TBG itself. Furthermore, some of these systems may have practical advantages over TBG in terms of the ease of fabrication or tunability of the physical properties.

A  distinguishing feature of the TBG physics is the appearance of remarkably flat bands at charge neutrality for magic twists $\theta_*$ \cite{Tarnopolsky2018,Bistritzer2011,Cao2018a}. The existence of such flat bands was predicted in Refs.~ \cite{Santos2007, Bistritzer2011} using an effective continuum model for two graphene layers with twist-independent interlayer coupling.  (see also  \cite{Santos2007, Bistritzer2011, Santos2012, Shallcross2010} and \cite{Li2010,Trambly2010,Santos2012}). In this model \cite{Bistritzer2011}, intra- and inter-sublattice hopping parameters were taken to be equal and band flattening happens at a certain sequence of magic angles for which the renormalized Fermi velocity vanishes at Dirac points. It was, however, recently realized that the effect of lattice relaxation in TBG leads to the expansion of the AB stacking regions relative to the AA regions in the Moir\'e pattern \cite{Carr2019}. As a result, the intra-sublattice hopping parameter $w_{\text{AA}}$ is suppressed relative to the inter-sublattice hopping parameter $w_{\text{AB}}$ at small twists. Crucially, this results in band gap opening and further band flattening, down to the point when the bands can in principle become absolutely flat \cite{Tarnopolsky2018}. Although it is understood that the existence of the flat bands is important for the correlated physics, it is still unclear which feature, - band-flattening, band isolation or band topology - is most decisive.

In this work, we report an infinite class of multilayer graphene systems which all manifest the remarkably flat bands and the corresponding magic angles patterns. 
Such systems, if realized experimentally, would provide a rich playground for correlated physics beyond TBG. It is worth noting that other multilayer systems studied in the literature such as ABC trilayer graphene stacked on hexagonal boron nitride  or twisted double bilayer graphene do not exhibit magic angles or flat bands when realistic effects, e.g. trigonal warping terms, are included despite the recent experimental observation of correlated insulting states and superconductivity \cite{TDBGexp2019, PabloTDBG, IOP_TDBG, Lee2019theory}.

\begin{figure}[t]
\includegraphics[width=0.49\columnwidth]{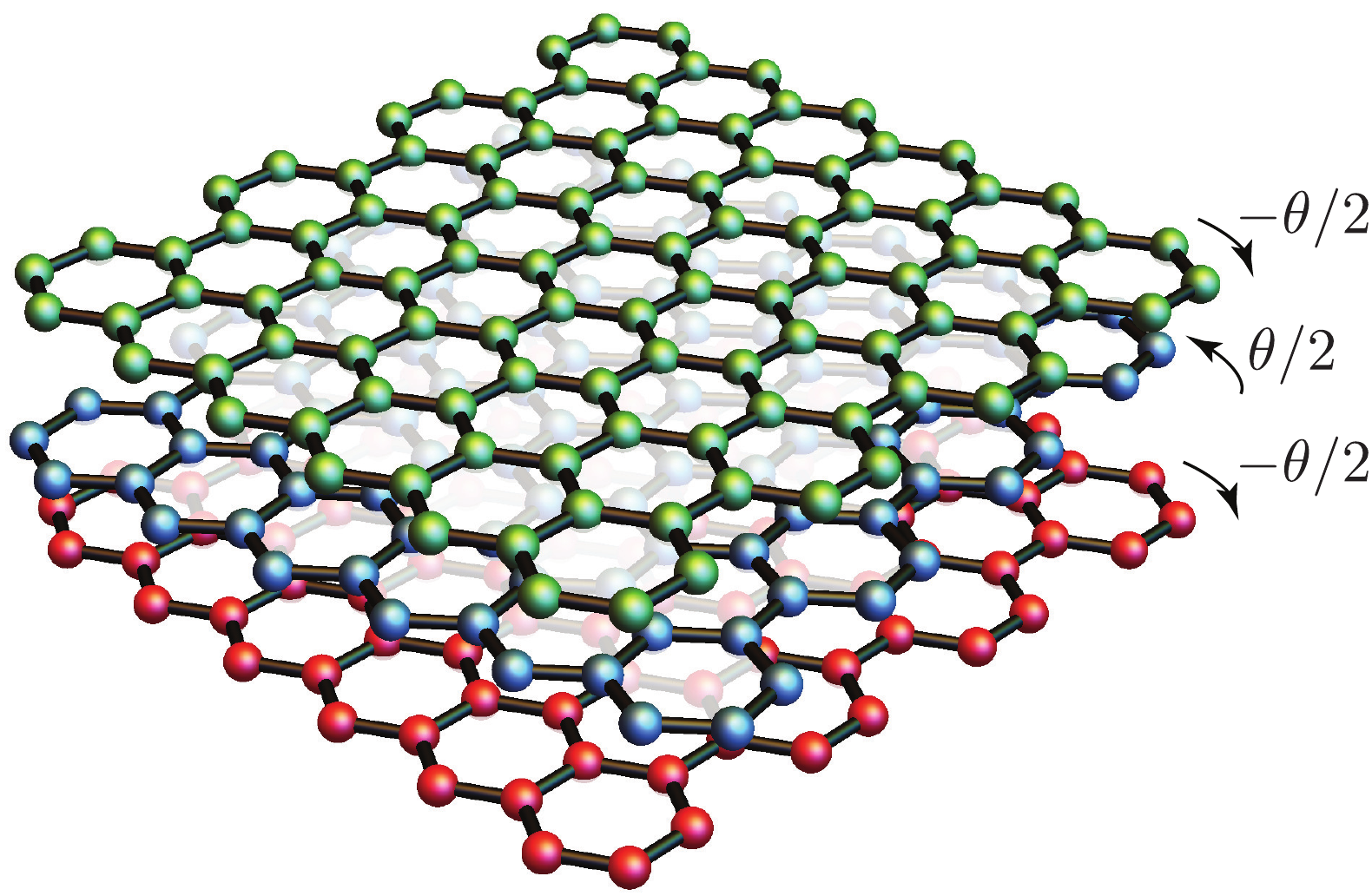}
\includegraphics[width=0.45\columnwidth]{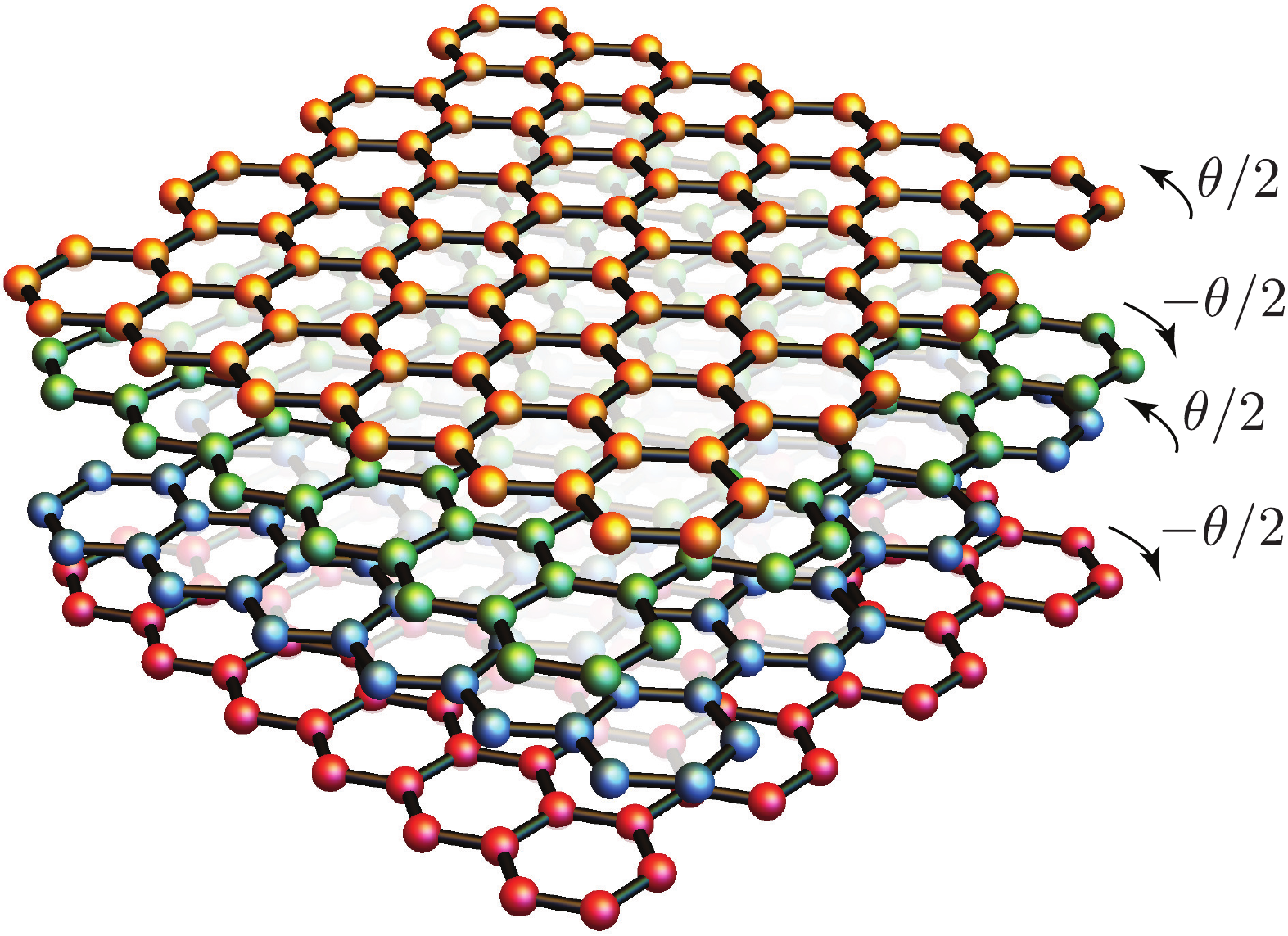}
\vspace{-3mm}
\caption{A schematic illustration of the alternating-twist multilayer graphene: trilayer (left) and quadrilayer (right). }
\label{Fig1}
\end{figure}

We consider a model of alternating-twist multilayer graphene (ATMG) for which the relative twists between two neighboring layers have the same magnitude but alter in sign (see Fig.\ref{Fig1}). In general, for a system with $n$ graphene sheets, there will be several Moir\'e patterns -- each one originating from a pair of adjacent layers. The overall periodicity of such patterns is determined by the relative inter-layer twists, while the origin of the pattern is controlled by the relative displacement. In the TBG case \cite{Bistritzer2011,Tarnopolsky2018} the resulting  Moir\'e physics is shift-independent, but in the case of a more complex multi-Moir{\'e} interference in twisted multilayers it can be relevant. In this work, we first focus on the case where the Moir{\'e} patterns are aligned so that the flat-band physics is most pronounced. At the end, we discuss how sensitive our results are to layer misalignment and show that the obtained flat bands are relatively stable to the inclusion of layer displacement.

Our main result is that the Hamiltonian for the ATMG with $n = 2n_e$ ($n = 2n_e + 1$) layers can be mapped {\it exactly} to a sum of $n_e$ twisted bilayer models (plus a single layer model) at {\it different} twist angles. Using this mapping, we find that there are $n_e$ sequences of magic angles given by multiplying the TBG magic angles by $2 \cos \frac{\pi k}{n+1}$ with $k=1, \dots, n_e$, which implies that the first magic angle for $n$-layered ATMG is larger than the first magic angle in TBG by a factor of $2 \cos \frac{\pi}{n + 1}$ which yields $\sqrt{2}$ for $n=3$ and approaches 2 as $n \rightarrow \infty$. This represents a practical advantage since samples with larger twist angles are generally more stable and easier to fabricate. In addition, the setup proposed here does not require independent tuning of the different relative twist angles since it can be achieved by repeatedly appying the "tear and stack" method.

\section{Model}

 We begin by considering a general system of $n$ graphene layers with the $\ell$-th layer twisted counter-clockwise around  {a} lattice cite by an angle $\theta_{\ell}$ and then displaced by a distance $\bd_{\ell}$ relative to a fixed reference. Similar to the bilayer problem, the coupling between layers $i$ and $j$ is characterized by two parameters $w_{\rm AA}^{ij}$ and $w_{\rm AB}^{ij}$ which indicate intra- and intersublattice coupling, respectively. We take these parameters to be generally different between different layers and assume coupling takes place only between nearest neighboring layers. The resulting low-energy effective Hamiltonian reads (see Appendix \ref{App:derivation} for details)
\beq
\label{H}
H =  \sum_{\ell=1}^n c_{\ell,\br}^\dagger   (-i v_F \bsigma_{\theta_{\ell}} \cdot \bnabla) c_{\ell,\br} + \sum_{\ell=1}^{n-1} c_{\ell,\br}^\dagger T_\br^{\ell,\ell+1} c_{\ell+1,\br} + \text{h.c.}, 
\eeq
where $v_F \approx 10^6$ m/s is the monolayer graphene Fermi velocity, $\bsigma_{\theta_{\ell}}=e^{\frac{i}{2}\theta_{\ell}\sigma_{z}}\bsigma e^{-\frac{i}{2}\theta_{\ell}\sigma_{z}}$ and the interlayer coupling matrix $T^{\ell,\ell+1}(\br)$ takes the form
\beq
\label{Tij}
T^{ij}(\br) = \left(\begin{array}{cc} 
 w^{ij}_{\textrm{AA}} U_0^{ij}(\br) &  w^{ij}_{\textrm{AB}} U_1^{ij}(\br) \\ 
 w^{ij}_{\textrm{AB}} U_1^{ij *}(-\br) & w^{ij}_{\textrm{AA}} U_0^{ij}(\br)
\end{array} \right),
\eeq
with the Moir\'e potentials $U_{m=0,1}^{ij}(\br)$ defined as
\beq
U^{ij}_m(\br) = \sum_{n=1}^3 e^{i m (n-1) \phi} e^{-i \bq_n^{ij} \cdot (\br - \bD_{ij})}\,.
\label{Uij}
\eeq

\noindent
Here, $\bq^{ij}_1 = 2 k_D \sin (\theta_{ji}/2) R_{\phi_{ij}}(0, -1)$, $\bq^{ij}_{2,3} = R_{\pm \phi} \bq^{ij}_1$, $\phi = 2\pi/3$, 
 $R_\theta = e^{-i \theta \sigma_y}$ denotes the counter-clockwise rotation operator  with angle $\theta$,  and $k_D = {4\pi}/{3 \sqrt{3} a}$ is the Dirac momentum of the monolayer graphene with lattice constant $a =1.42$ \AA. We also introduced the auxiliary angle variables $\theta_{ji} = \theta_j - \theta_i$ and $\phi_{ij} = (\theta_i + \theta_j)/2$. We can write the displacement vector $\bD_{ij}$ of the Moir\'{e} pattern as
\beq
\bD_{ij} = \frac{\bd_i + \bd_j}{2} + i \cot (\theta_{ji}/2) \sigma_y \frac{\bd_i - \bd_j}{2}.
\label{Dij}
\eeq

In the bilayer case ($n=2$), the Hamiltonian (\ref{H}) reduces to the TBG Hamiltonian \cite{Santos2007,Bistritzer2011,Tarnopolsky2018} up to the gauge transformation $c_{{\ell}} \rightarrow c_{{\ell}} e^{i R_{\theta_{{\ell}}} K \cdot \bd_{{\ell}}}$. The advantage of the form we consider here is that it makes it clear how the layer displacements $\bd_i$ enter the Hamiltonian by shifting the corresponding Moir\'e potentials. For $n$ layers, there are $n-1$ shift variables $\bD_{\ell,\ell+1}$, $\ell=1,\dots,n-1$, one of which can be removed by redefining the origin, leaving $n-2$ variables which influence the spectrum. This is the reason why the shift vectors were unimportant in the bilayer case in contrast to the multilayer case considered here.

In general, the potential (\ref{Tij}) will generate several overlapping Moir\'{e} patterns generated by the different angles and shift vectors between consecutive layers. For most of this paper, we will focus on the case of "unshifted" ATMG which corresponds to the choice $\theta_{\ell}= (-1)^{\ell} \theta/2$ and $\bd_{\ell} = \bd$ such that the nearest neighboring layers are aligned and have alternating relative twists of $\pm \theta$. In this case, $\phi_{\ell,\ell+1} = 0$, $\theta_{\ell+1,\ell} = (-1)^{\ell+1} \theta$, $\bD_{\ell,\ell+1} = \bd$ and there is a single Moir\'e pattern similar to the bilayer problem. We also assume that the ratio between $w_{\textrm{AA}}^{ij}$  and $w_{\textrm{AB}}^{ij}$ couplings is layer-independent and denote it as 
\beq
\label{kappadef}
\kappa = w_{\textrm{AA}}^{ij}/w_{\textrm{AB}}^{ij}\,.
\eeq
Assuming small twist angle $\theta$, we can neglect the phase factor in the Pauli matrices $\bsigma_{\theta_{\ell}}\to \bsigma$  and get rid of the angular dependence by introducing the dimensionless variables $\alpha_{ij} = w_{\textrm{AB}}^{ij}/(v_F k_D \theta)$  leading to\footnote{This assumption is unnecessary in the chiral limit ($\kappa = 0$) since the phase can be removed by a gauge transformation.}
\beq
\label{HDM}
\H = \left(\begin{array}{cc} \M & \D^\dagger \\ \D & \M \end{array} \right)_{\rm AB}\,,
\eeq
where AB indicates the matrix is in the sublattice space. The operators $\D$ and $\M$ are given by
\begin{gather}
\D = \left(\begin{array}{cc} -2 i \bar \partial & W U_1(\br) \\ W^T U_{1}(-\br)  & -2 i \bar \partial \end{array}\right), \\ \M = \kappa \left(\begin{array}{cc} 0 &  W U_0(\br) \\  W^T U_0(-\br) & 0 \end{array}\right)\,.
\end{gather}
Here, we have rescaled the Hamiltonian so that all energies are measured in units of $v_F k_D \theta$. We also rescaled the coordinates so that they are measured in terms of the Moir\'e length scale $\br \rightarrow k_D \theta \br$ and introduced the derivatives $\partial$ and $\bar \partial$ relative to the dimensionless complex variable $z = x + i y$. The potentials $U_{m}(\br)$ are given by (\ref{Uij}) with $\bD_{ij}=0$  and $\bq^{ij}_n=R_{(n-1)\phi} (0, -1)$. The operators $\D$ and $\M$ act on vectors which has the form $\psi = (\psi_o, \psi_e)^T$ where $\psi_{o/e}$ contain the wave functions for the odd/even layers given explicitly as $\psi_o = (\psi_1, \psi_3, \dots, \psi_{2n_o-1})^T$ and $\psi_e = (\psi_2, \psi_4, \dots, \psi_{2 n_e})^T$ where $n_e = \lfloor n/2 \rfloor$ and $n_o = \lceil n/2 \rceil$ {are numbers of even and odd layers}. The matrix $W$ is  $n_o \times n_e$ layer hopping matrix and  given by
\beq
W = \left(\begin{array}{cccc} 
\alpha_{12} & 0 & 0 & \dots \\
\alpha_{23} & \alpha_{34} & 0 & \dots \\
0 & \alpha_{45} & \alpha_{56} & \dots \\
\dots & \dots & \dots & \dots 
\end{array}\right)\,.
\eeq

\section{Results}
\subsection{Reduction to the bilayer problem}
We now show that the Hamiltonian of the multilayer problem with $n = 2{n_{e}}$ ($n = {2n_{e}+1}$) layers can be mapped {\it exactly} to a direct sum of ${n_{e}} $ bilayer Hamiltonians (plus a  single layer Hamiltonian). This is done by writing the singular value decomposition of $W$ as $W = A \Lambda B^\dagger$ where $A$ and $B$ are $n_o \times n_o$ and $n_e \times n_e$ unitary matrices, respectively, and $\Lambda$ is an $n_o \times n_e$ matrix with $\lambda_{k}$, $k=1,\dots, n_e$ on the diagonal and  zeros everywhere else ($\lambda_{k}$ are  square roots of the eigenvalues of $W^{T} W$). Applying the unitary transformation given by $V = \diag(A, B)$ in the odd/even space to the Hamiltonian (\ref{HDM}) yields
\beq
V^\dagger \H V = \begin{cases}
\H^{(2)}_{\lambda_1} \oplus \H^{(2)}_{\lambda_2} \dots \oplus \H^{(2)}_{ \lambda_{n_e}}, & \text{$n$ even}, \\
\H^{(2)}_{\lambda_1}  \oplus  \H^{(2)}_{\lambda_2} \dots \oplus \H^{(2)}_{\lambda_{n_e}} \oplus \H^{(1)}, & \text{$n$ odd}, \\
\end{cases}
\label{VHV}
\eeq
where $\H^{(2)}_{\alpha}$ is the bilayer Hamiltonian with coupling parameters $\alpha = w_{\textrm{AB}}/(v_F k_D \theta)$ and $\kappa $ and $\H^{(1)}$ is the Hamiltonian for a single graphene layer. 

A consequence of the preceding discussion is that the spectrum of the multilayer problem with coupling matrix $W$ is given by the union of the spectra of several bilayer problems whose coupling parameters are given by the eigenvalues of the matrix $\sqrt{W^T W}$ (in addition to a single layer graphene dispersion if the number of layers is odd). Moreover, the eigenstates of the multilayer problem are easily obtainable from the eigenstates of the single layer problem. This applies particularly for the case of flat bands where the eigenstates were shown to have a simple form \cite{Tarnopolsky2018}.

\subsection{The chiral limit}  The chiral model for twisted bilayer graphene where the same-sublattice coupling set to zero $w_\text{AA} =0$ (or equivalently $\kappa = 0$ in this work) was introduced in Ref.~\cite{Tarnopolsky2018}. It was shown that this model captures the essential phenomenology of magic angles where the different notions of flatness (vanishing Fermi velocity, minimum bandwidth, maximum band gap) all coincide due to the appearance of perfectly flat bands for special (magic) values of the dimensionless coupling $\alpha = w_{\text{AB}}/(v_F k_D \theta)$. This model is one of the simplest models exhibiting magic angle flat bands and its applicability to TBG is supported by the observation that lattice relaxation tends to reduce the size of AA regions relative to AB regions \cite{Carr2019}, thus suppressing the value of $w_\text{AA}$ (intrasublattice coupling) relative to $w_\text{AB}$ (intersublattice coupling). 

\begin{figure}[t]
\center
\bgroup
\setlength{\tabcolsep}{0.3 em}
\begin{tabular}{c|c|c|c|c|c|c}
\hline
$n$ & $\alpha_1$ & $\alpha_2$ & $\alpha_3$ & $\alpha_4$ & $\alpha_5$ & $\alpha_6$\\
\hline
2 & 0.586 & 2.221 & 3.75 & 5.276 & 6.795 & 8.313\\
\hline
3 & 0.414 & 2.57 & 2.652 & 3.731 & 4.805 & 5.878\\
\hline
\multirow{2}{*}{4}  & 0.362 & 1.372 & 2.318 & 3.261 & 4.2 & 5.138 \\
 & 0.948 & 3.594 & 6.067 & 8.537 & 10.995 & 13.4507 \\
 \hline
 \multirow{2}{*}{5}  & 0.338 & 1.282 & 2.165 & 3.046 & 3.923 & 4.8 \\
 & 0.586 & 2.221 & 3.75 & 5.276 & 6.795 & 8.313\\
 \hline
 \multirow{3}{*}{6}  & 0.325 & 1.233 & 2.081 & 2.928 & 3.771 & 4.613 \\
 & 0.47 & 1.781 & 3.007 & 4.231 & 5.45 & 6.667\\
 & 1.317 & 4.99 & 8.426 & 11.855 & 15.268 & 18.679 \\
 \hline
\end{tabular}
\egroup\\
\vspace{0.2 cm}
\includegraphics[width= 0.9 \columnwidth]{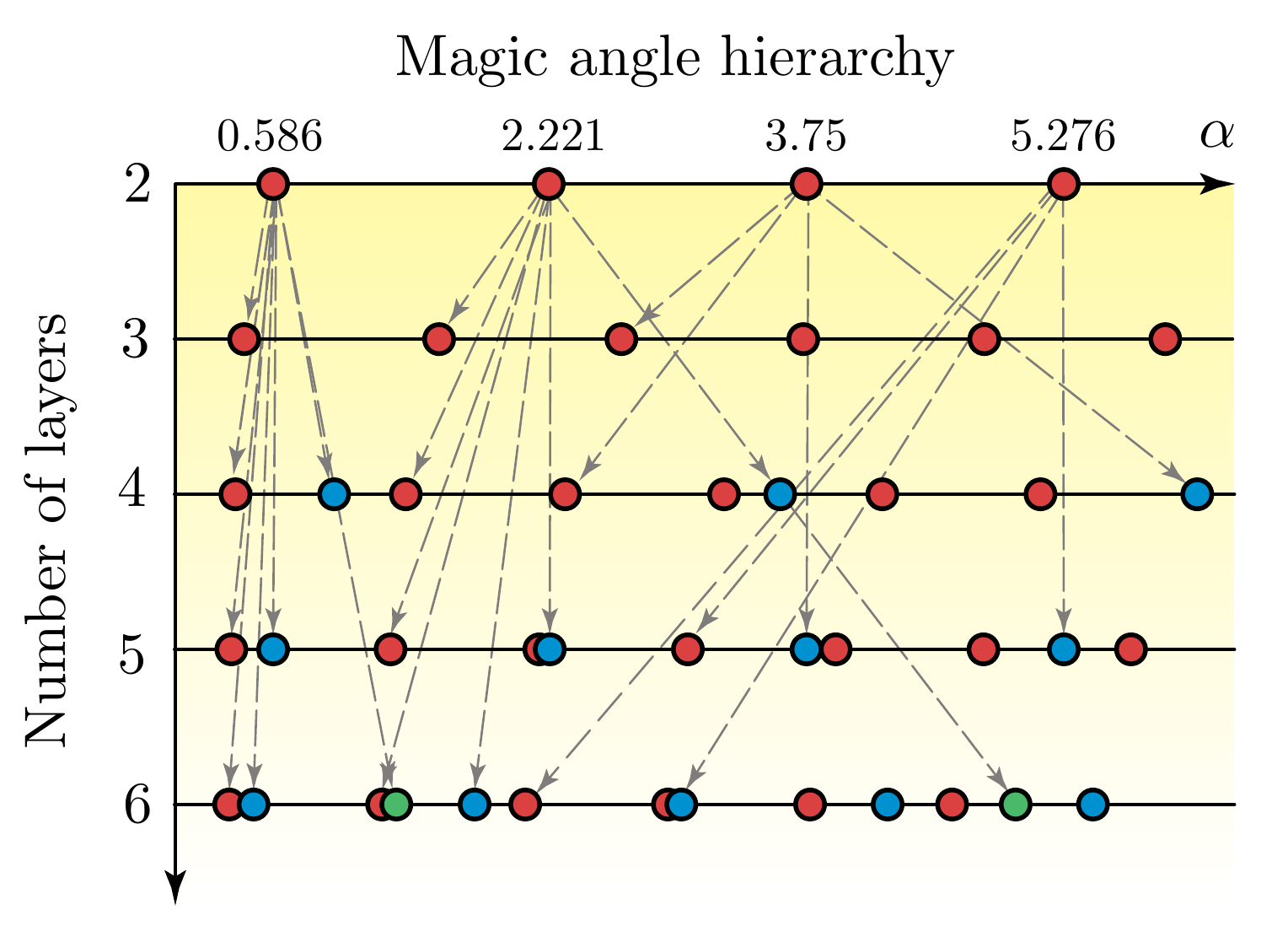}
\caption{Magic angles families for the alternating-twist multilayer graphene in the chiral limit ($\kappa=0$). Upper panel: Numerical values of magic angles for alternating-twist $n$-layered systems ($n=2,3..$). The magic angle parameters $\alpha_i = w_{\text{AB}}/(v_F k_D \theta_i)$ designate the twists under which the lowest bands become perfectly flat. For each $n$, there are $\lfloor n/2 \rfloor$ sequences of magic angles (denoted by different colors) obtained by dividing the bilayer magic angles by $2 \cos \frac{\pi k}{n+1}$, $k = 1,\dots,\lfloor n/2 \rfloor$ as illustrated schematically in the lower panel.}
\label{MA}
\end{figure}

Let us first consider the standard setting where all interlayer couplings are the same $\alpha^{ij} = \alpha$. In this case, the layer hopping $n_o \times n_e$ matrix $W $ is given by
\beq
W = \alpha(\delta_{i j} + \delta_{i,j+1})\,.
\label{MM}
\eeq
The eigenvalues of $\sqrt{W^T W}$ can be easily computed for any number of layers $n$ and they are given by $\lambda_k = 2 \cos (\frac{\pi k}{n+1} )\alpha $, $k = 1,\dots, n_e$. Thus, the ATMG with $n$ layers has $n_e$ sequences of magic angles given by 
\beq
\alpha^{(n)}_k = \alpha^{(2)}/\big(2 \cos \frac{\pi k}{n+1}\big)\,.
\eeq
Here, $\alpha^{(2)}$ is the sequence of magic angles in the bilayer problem which was computed in \cite{Tarnopolsky2018} as $\alpha^{(2)} = 0.586, 2.221, 3.75, 5.276, 6.795, \dots$. The magic angle sequence for any $n$ can then be easily computed as shown in Fig.~\ref{MA} for $n$ up to 6.

It should be noted that the mapping to the bilayer problem can also be used to explicitly write the wave functions of the multilayer system in terms of their bilayer counterparts. For the eigenvalue $\lambda_k$, the corresponding eigenfunction{s} for all layers $\ell =1,\dots, n $ {are}  given by
\beq
\psi_{\ell}^{(k)}(\textbf{r}) =  \sqrt{\frac{2n}{n+1}} (-1)^{\ell(k+1)}\sin(\frac{\pi k}{n+1}\ell)\psi^{\textrm{TBG}}_{{\lambda_{k}}}(\textbf{r})\,,
\eeq
normalized as $\sum_{\ell=1}^{n}|\psi_{\ell}^{(k)}(\textbf{r}) |^{2}= n|\psi^{\textrm{TBG}}_{{\lambda_{k}}}(\textbf{r})|^{2}$  {and 
$\psi^{\textrm{TBG}}_{\lambda_{k}}(\textbf{r})$ is the eigenfunction of the TBG Hamiltonian $\mathcal{H}^{(2)}_{\lambda_{k}}$}.

\begin{figure}
\center
\includegraphics[width=0.48\textwidth]{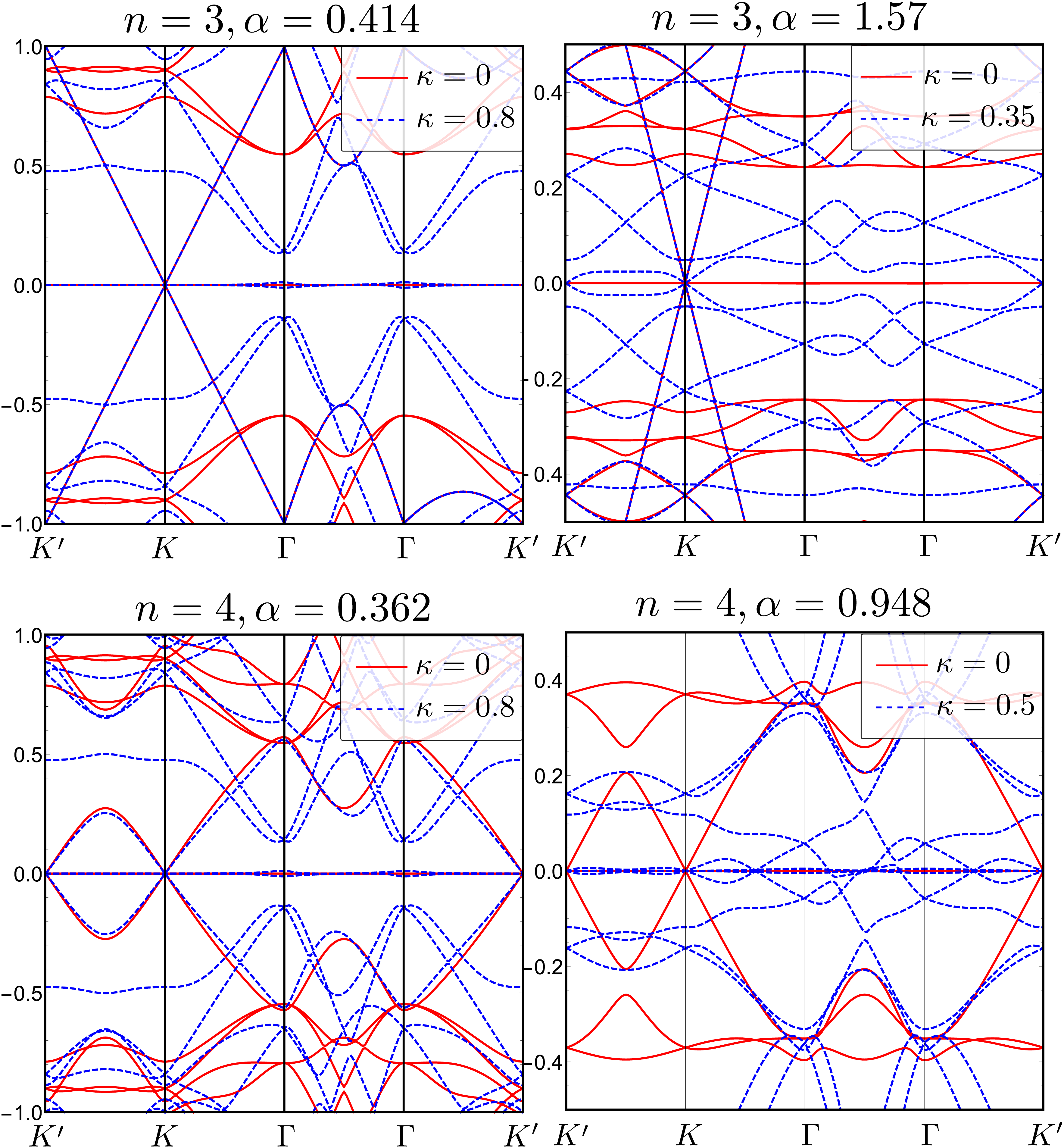}
\caption{Band structure at the first two magic angles for the trilayer $n=3$ and quadrilayer $n=4$ cases. We show the spectrum for the chiral limit $\kappa = 0$ (red, solid) as well as the realistic lattice relaxation value for $\kappa$ at the corresponding angle \cite{Carr2019} (blue, dashed). In the chiral limit, we can observe a perfectly flat band coexisting with a single Dirac cone at the $K$ point for $n=3$. For $n=4$, the chiral flat band coexists with another tBG spectrum at non-magic angle. We can see that the flat bands for the first magic angle for $n=3$ (upper left) and the first two magic angles for $n=4$ (lower left and right) are stable to the addition of intrasublattice interlayer coupling $\kappa \neq 0$, whereas the flat band at the second magic for $n=3$ (upper right) gets destroyed. All energies are measured in units of $\hbar v_F k_D \theta = w_{\rm AB}/\alpha$.}
\label{FB34}
\end{figure}

For the trilayer case ($n=3$), the only eigenvalue of $\sqrt{W^T W}$ is $\lambda_1 = \sqrt{2}\alpha $. Eq.~\ref{VHV} implies that the system is equivalent to the sum of a bilayer problem with coupling $\sqrt{2} \alpha$ and a single layer problem. As a result, we can immediately read off the magic angles, where a perfectly flat band appears, to be $\alpha^{(3)} = \alpha^{(2)}/\sqrt{2} = 0.414, 1.57, 2.65,3.731, 4.805, \dots$. This is verified in Fig.~\ref{FB34}, where the band structure is computed numerically for the first two magic angles for the trilayer problem showing the existence of a perfectly flat band. 

One particularly interesting feature here is that for the trilayer graphene, the first magic angle is larger by a factor of $\sqrt{2}$ compared to the TBG, which represents an experimental advantage. Apart from the scaling of the magic angles, the trilayer system differs from TBG in two main aspects. First, the flat band here coexists with a dispersing Dirac cone.  If realized experimentally, this feature will distinguish the physics of the trilayer system  from the TBG physics, and could help to elucidate whether \textit{band flatness} or \textit{band isolation} plays the bigger role in the correlated physics. Second, while the band structure for the flat band looks identical to the TBG band structure at the first magic angle $\theta \approx 1.08^o$, the actual scale for the Moir\'e pattern is determined by the actual angle $\theta \approx 1.53^o$ which determines the scale of the gaps and the interaction. We stress that the mapping does not only apply for the spectra but also for the wave functions. As a result, the physics of the trilayer model (including the interaction effects) will be identical to the physics of TBG with all distances scaled down by a factor of $\sqrt{2}$ and with an extra Dirac band from an individual graphene layer.

For the quadrilayer case ($n=4$), the matrix $\sqrt{W^T W}$ has two eigenvalues $\lambda_{1,2}=\alpha \varphi^{\pm 1}$, where $\varphi = (\sqrt{5} + 1)/2$ is  the golden ratio, yielding two sequences of magic angles $\alpha^{(4)} = \alpha^{(2)}/\varphi = 0.362, 1.373, 2.318, 3.261, 4.2, 5.138, 5.075, \dots$ and $\alpha'^{(4)} = \alpha^{(2)} \varphi = 0.948, 3.594, 6.069, 8.537, \dots$. The quadrilayer ATMG maps to a sum of two TBGs and these two sequences correspond to points at which one of these two twisted bilayers hits a magic angle. The largest magic angle (the smallest $\alpha$) in this case is $\theta \approx 1.75^\circ$ -- which is larger than the bilayer and trilayer cases. 

\begin{figure}
\center
\includegraphics[width=0.45\textwidth]{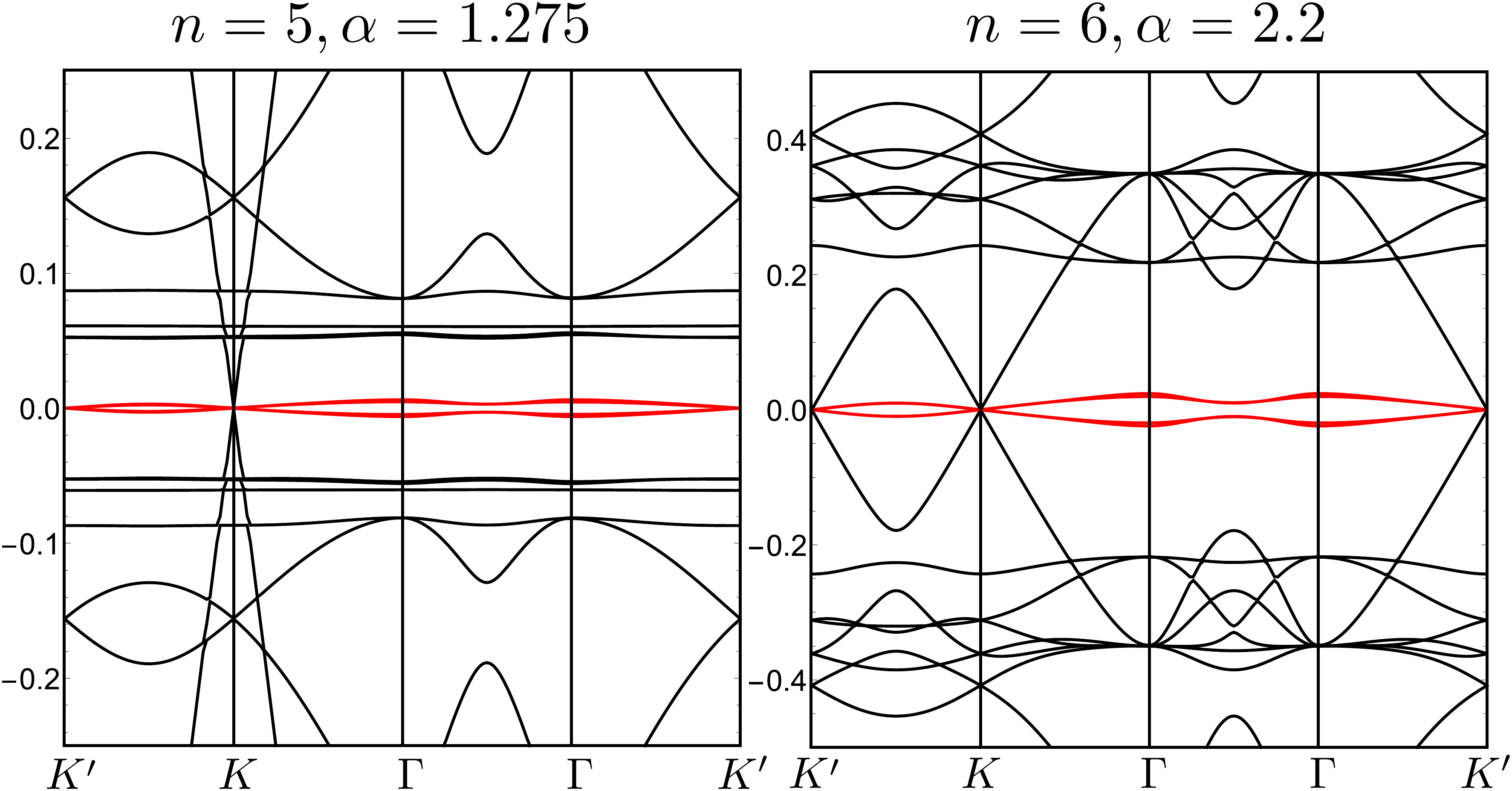}
\caption{Band structure for the models with $n=5$ layers with $\alpha = 2.2$ and $n=6$ layers with $\alpha = 1.275$. In both cases, the vicinity to two very close magic angles leads to the appearance of two pairs of almost perfectly flat bands. All energies are measured in units of $\hbar v_F k_D \theta = w_{\rm AB}/\alpha$.}
\label{ADFB}
\end{figure}

We note that since the different magic angle sequences for a given $n$ have incommensurate periods, we can find some values of $\alpha$ which is close to several magic angles from different sequences simultaneously. This happens for example for $n=5$ for $\alpha \approx 2.2$ which is very close to the third magic angle in the first sequence (2.165) and the {second} magic angle in the second sequence (2.221). Another example happens when $n=6$ and $\alpha \approx 1.275$ which is very close to the second magic angle in the first sequence (1.23) and the first magic angle in the third sequence (1.32). In both cases, there are two pairs of very narrow bands coexisting at $0$ as shown in Fig.~\ref{ADFB}.

It is instructive to consider the limit of large number of layers $n \rightarrow \infty$. {In this case, the eigenvalues  $\lambda_k = 2 \cos (\frac{\pi k}{n+1}) \alpha$ of the matrix $\sqrt{W^T W}$  form a continuum from $0$ to $2$}, which implies that there is a continuum of magic angles: whenever $\alpha > \alpha^{(2)}_1/2 \approx 0.293$, we are always arbitrarily close to a magic angle descending from the first magic angle of TBG where at least a single band is perfectly flat band. Similarly, there is a flat band deriving from the second magic angle for $\alpha > \alpha^{(2)}_2/2 \approx 1.11$. In general, there will be exactly $k$ perfectly flat bands deriving from the first $m$ TBG magic angles whenever $\alpha^{(2)}_{m-1}/2 \leq \alpha < \alpha^{(2)}_{m}/2$, where $\alpha^{(2)}_m$ is the $m$-th magic angle of TBG.  This suggests an intriguing connection to possible flat-band-related phenomena in some samples of turbostratic graphites,  if its layers are naturally assembled in small but very random alternating twists \cite{Shallcross2010}. 

When the number of layers $n$ is relatively small, it is still possible to achieve several flat bands at $0$ simultaneously if we allow for different hopping parameters between different layers. So far, we have only considered the case where all the couplings $\alpha_{ij}=\alpha$ are equal which is naturally expected since all the graphene layers are identical. We now  consider instead the possibility that the coupling between layers is non-uniform. For instance, the coupling to the outer layers (top and bottom) may differ slightly from that between inner layers. Another possibility is to artificially tune the couplings by including thin layers of a dielectric material between some of the layers or by depositing adatoms on the top or bottom surfaces to change the interlayer potential. Our purpose in this discussion is to show that this is an interesting theoretical possibility leaving the question of experimental realizability to future studies.

\begin{figure}
\includegraphics[width=0.48\textwidth]{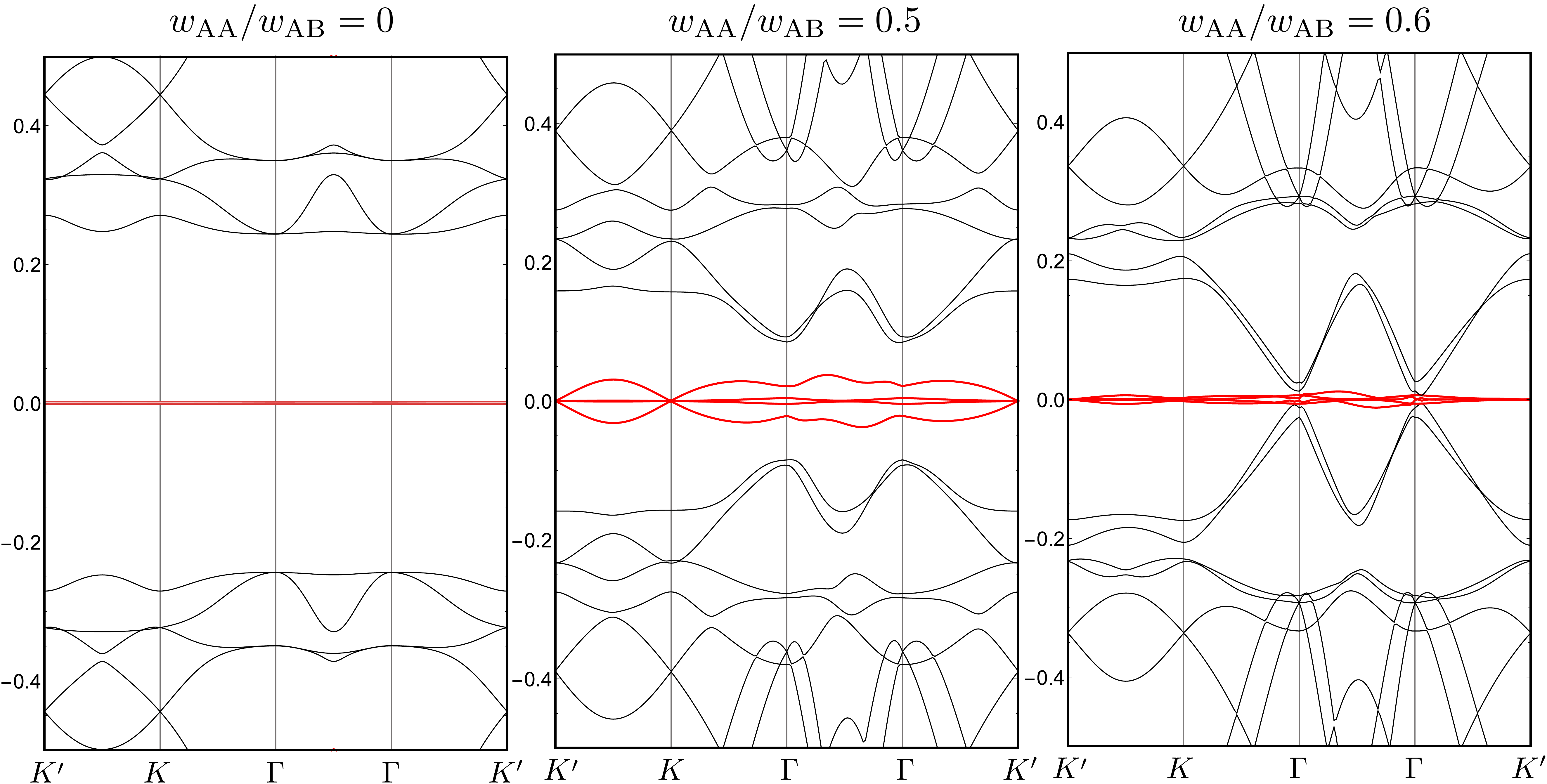}
\caption{Band structure for the quadrilayer problem with unequal layer couplings $\alpha_{12} = \alpha_{34} = 1.14$, $\alpha_{23} = 1.64$. We notice the appearance of a pair of perfectly flat bands which are relatively stable to the inclusion of intrasublattice intralayer coupling $\kappa=w_{\text{AA}}/w_{\text{AB}}$. All energies here are measured in units of $\hbar v_F k_D \theta$}
\label{DFB}
\end{figure}

Let us now consider the simplest case with four layers $n=4$ such that the coupling to the outer layers $\alpha_{12} = \alpha_{34} = \alpha_1$ is different from the coupling between the middle layer $\alpha_{23} = \alpha_2$. In this case, we can achieve two perfectly flat bands simultaneously as follows: we require the two eigenvalues of the matrix $\sqrt{W^T W}$ to be equal to the first two magic angles. A simple way to achieve this is to require the determinant and the trace of this matrix to be equal to the {product} and sum of the first two bilayer magic angles $\alpha^{(2)}_{1,2}$ leading to the equations
\beq
\alpha_1^2 = \alpha^{(2)}_1 \alpha^{(2)}_2, \qquad 2\alpha_1^2 + \alpha_2^2 = (\alpha^{(2)}_1)^2 + (\alpha^{(2)}_2)^2\,,
\eeq
which can be easily solved for $\alpha_{1,2}$ yielding $\alpha_1 = \sqrt{\alpha^{(2)}_1 \alpha^{(2)}_2} = 1.14$ and $\alpha_2 = |\alpha^{(2)}_1 - \alpha^{(2)}_2| = 1.6{4}$. The band structure for this choice of parameters is shown in Fig.~\ref{DFB} showing two perfectly flat bands.

\subsection{Switching on intrasublattice couplings $w_{\text{AA}}$} We now consider non-zero intra-sublattice coupling $w_{\text{AA}}$  which in our notation corresponds to non-zero $\kappa$ (cf.~Eq.~\ref{kappadef}). The value of $\kappa$ generally depends on the angle; it is expected to be close to 1 for $\alpha \lesssim 0.25$ then starts decreasing as $\alpha$ is increased \cite{Carr2019}. In the following, we will use the values of the relaxation parameter $\kappa$ computed for TBG \cite{Carr2019} as an estimate for the multilayer problem although a more involved \textit{ab-initio} calculation is needed to refine this value and account for complex lattice relaxation effects in multilayer systems.

In TBG, it is known that the flat bands at the first magic angle are a lot more stable than higher magic angles when $\kappa$ is non-zero \cite{Tarnopolsky2018}. Our mapping implies that the flat bands in the multilayer model inherit the stability of the corresponding bilayer flat bands. That is, multilayer magic angles which descend from the first bilayer  magic angle are significantly more stable than those descendent from higher magic angles. 

For the trilayer case ($n=3$), this means that the first magic angle $\alpha = 0.414$ is stable for relatively large values of $\kappa$ including the realistic value $\kappa \approx 0.7-0.8$ estimated in Ref.~\cite{Carr2019}. The second magic angle $\alpha = 1.57$ is however significantly less stable and the flat band gets destroyed for realistic values of $\kappa \approx 0.4$. For quadrilayer ($n=4$), the situation is different since the first two magic angles $\alpha = 0.362, 0.948$ are descendent from the first magic angle of TBG. In fact, the flat band $\alpha = 0.948$ inherits the stability of the bilayer flat band while also having a significantly larger value of relaxation $\kappa \approx 0.5$ compared to the second magic angle in TBG which leads to extra stability. The flat bands for the first two magic angles for $n=3, 4$ are shown in Fig.~\ref{FB34} for realistic values of the relaxation parameter $\kappa$ and we can see that they are stable in all cases except the second magic angle for $n=3$ as expected. The effect of $\kappa$ on the $n=4$ setup with different couplings leading to two perfectly flat bands can also be investigated and we find that it is relatively stable for realistic values of $\kappa$ around $0.5-0.6$ (cf.~Fig.~\ref{DFB}). 
  
 \section{Experimental realization}
So far we have focused on the setting where the layers are perfectly aligned, i.e. the displacement vectors $\bD_{l,l+1}$ defined in (\ref{Dij}) were assumed to be equal. For experimental realizations, it however very difficult to achieve perfect alignment on the atomic scale. Hence, it is crucial to investigate how our results are affected when we lift this assumption and consider unequal displacement vectors. In this case, the exact mapping to TBG no longer exists. However, the Hamiltonian (\ref{H}) is still translationally invariant on the Moir\'e lattice (since the different Moir\'e potentials are only shifted relative to one another) and we can find the band structure within the Moir\'e Brillouin zone numerically. 
 
\begin{figure}[t]
\center
\includegraphics[width=0.5\textwidth]{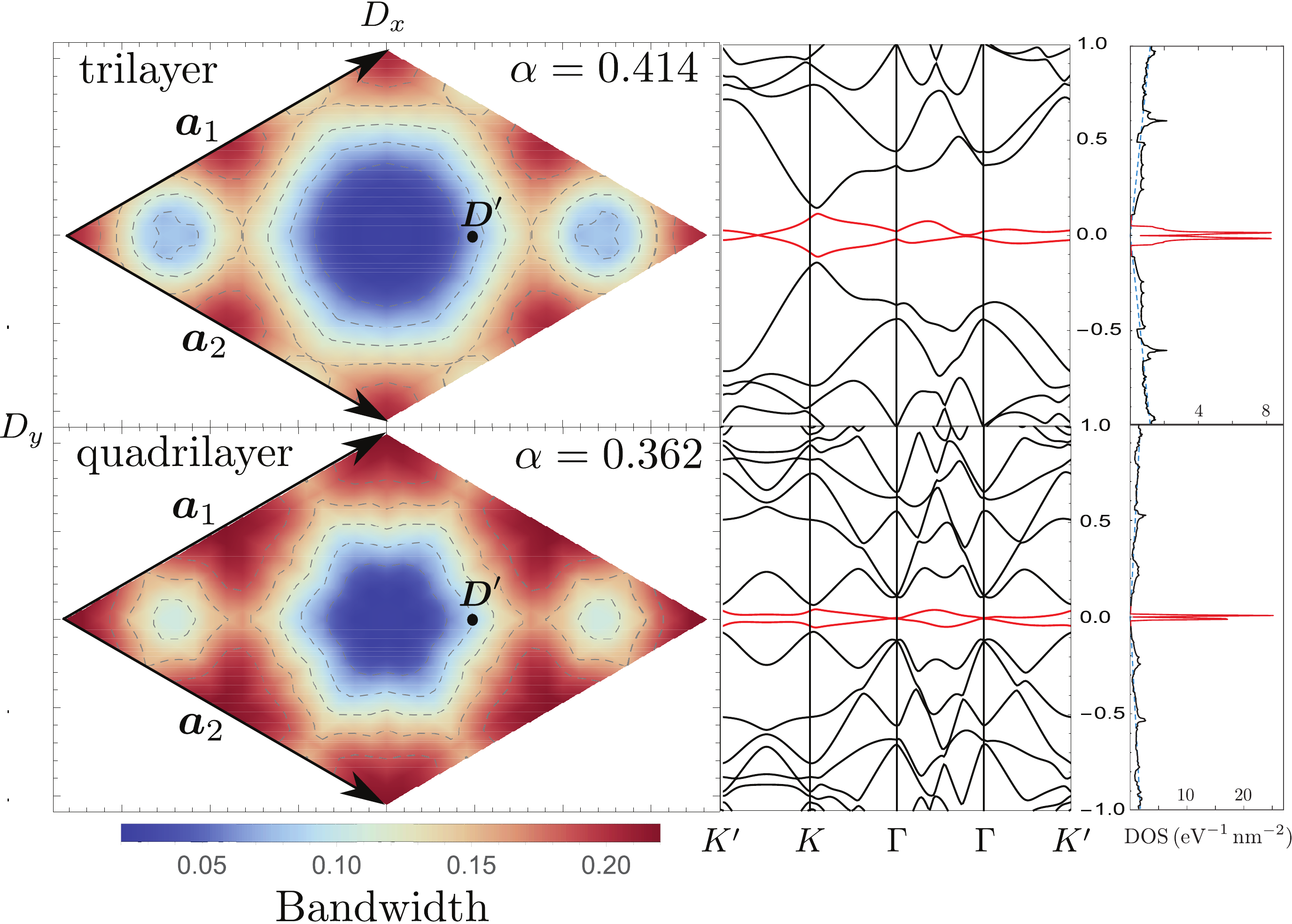}
\caption{Band width in units of $\hbar v_F k_D \theta$ as a function of the displacement vector $\bD$ within the Moir\'e unit cell formed by the vectors $\ba_{1,2} = \frac{4\pi}{3}(\frac{\sqrt{3}}{2}, \pm \frac{1}{2})$ for $n=3$ and $n=4$ at the first magic angle for $\kappa = 0.8$ (left panel) together with the band structures for one selected point $\bD' = (1,0)$ at the border of the blue (narrow band) region (middle panel). The density of states exhibits a very large split peak close to charge neutrality due to the flat band (left panel).}
\label{BW}
\end{figure}

 For the trilayer case ($n=3$), we can set the displacement $\bD_{12}$ to zero by shifting the origin leaving one relevant displacement $\bD_{23} = \bD$. The bandwidth of the narrow band for $\kappa = 0.8$ and for different values of the shift vector $\bD$ is shown in Fig.~\ref{BW}. We can see that there is a range of $\bD$ around $0$ for which the bandwidth remains relatively small. For a reference, we can compare the bandwidth with the Coulomb energy scale given by ${E_{\textrm{Coulomb}}} =  \frac{e^2 \theta}{4 \pi \epsilon \epsilon_0 a}$. In our dimensionless unit, $\epsilon_{\textrm{Coulomb}} = \frac{E_{\textrm{Coulomb}}}{\hbar v_F k_D \theta} \approx \frac{1}{\epsilon_0}$. For $\epsilon_0 \approx 4$, this scale exceeds the bandwidth for the whole range of $\bD$ as shown in Fig.~\ref{BW} implying strong interaction effects regardless of the layer displacement. We notice that, even when the band is not perfectly flat, it is associated with a very large peak in the density of states as shown in right panel of Fig.~\ref{BW}.
 
 For the quadrilayer case ($n=4$), we can set $\bD_{23}$ to zero leaving two different shift vectors $\bD_{12}$ and $\bD_{34}$ which affect the spectrum. However, if the quadrilayer system is made using the tear and stack method from the same TBG, then the displacements satisfy $\bd_4 - \bd_3 = \bd_2 - \bd_1$. For small enough angles, the displacement vectors are $\bD_{12} = i \sigma_y (\bd_1 - \bd_2)/\theta$ and $\bD_{34} = i \sigma_y (\bd_3 - \bd_4)/\theta = \bD_{12}$ \footnote{Notice that the layer displacement $\bd_{\ell}$ is defined up to translations of the original lattice whereas the effective shifts $\bD_{\ell,\ell+1}$ are defined up to the Moir\'e lattice translations.}, hence there is only one shift vector $\bD = \bD_{12} = \bD_{34}$ and we can again investigate how it influences the bandwidth as shown in Fig.~\ref{BW}. Similar to the trilayer case, we see that the bandwidth of the lowest band is small compared to the interaction scale for the whole range of displacements. We also see in this case that the narrow band is associated with a very large peak in the density of states.
 
 It follows from the previous discussion that perfect alignment of the layers is not a requirement for the appearance of flat bands in ATMG since the bandwidth remains reasonably small for the whole range of displacements. It is worth noting that by using the "tear and stack" trick repeatedly, we can ensure that the twist angles between consecutive layers are exactly equal and opposite without the need of any extra tuning for the angles. For trilayers, the procedure would start by tearing a part of a monolayer sample, twisting and stacking it, then tearing another piece of the base monolayer sample and stacking it on the top of the twisted bilayer without any extra twisting. This ensures that the top and bottom layers are aligned which implies that the two twist angles are opposite to a very good accuracy. The same can be done for quadrilayer by using the tear and stack method starting with a TBG sample, again without any additional twisting. Thus, we expect it to be possible to realize alternating twist angles of equal magnitude to a reasonable accuracy. Adding the fact that the magic angles in the multilayer setting are larger and thus more stable makes our current setting a very promising setup to observe magic angle physics beyond TBG.

\section{Conclusion}
In conclusion, we have introduced a model of twisted multilayer graphene with alternating twist angle focusing on the limit of aligned layers. We have shown that this model for $n=2{n_{e}}$ ($n=2{n_{e}}+1$) layers maps exactly to a sum of ${n_{e}}$ twisted bilayer models (plus a single layer model) with different twist angles. Such mapping enabled us to determine the pattern of magic angles for arbitrary $n$ which is given by multiplying the bilayer magic angles by $ 2 \cos \frac{\pi k}{n+1}$, $k = 1,\dots,n_e$. Focusing on the trilayer and quadrilayer cases, we found that these models exhibit flat bands coexisting with other dispersing bands at zero energy and showed that such flat bands are relatively stable even when layer misalignment is taken into account. In addition, we found that for relatively large number of layers or when interlayer couplings are different, we can achieve several flat bands at zero energy simultaneously. Moreover, we show that there is a continuum of magic angles for $\theta \lesssim 2^{{\circ}}$ in the limit of very large number of layers $n \rightarrow \infty$. This might suggest an intriguing link to possible flat-band-related phenomena in some samples of turbostratic graphites,  if its layers are  naturally assembled in small but very random alternating twists. At the end, we discuss possible experimental realizations of the model and show that it can be achieved within current technology by applying the tear and stack method repeatedly. Compared to TBG, the multilayer setting has the advantage that the first magic angle is larger making it easier to realize experimentally.

\

\noindent {\bf Note}: During the preparation of this manuscript, two related preprints appeared: \cite{Mora19} which discussed twisted trilayer graphene with commensurate twist angles  and \cite{Cea19} which discusses several multilayer settings including a related one with equal rather than alternating twist angles.


\

\acknowledgements
 We thank Bertrand Halperin, Philip Kim and Pablo Jarillo-Herrero for fruitful discussions. A.J.K. was supported by the Swiss National Science Foundation’s grant
P2ELP2\_175278. G.T. was supported by the MURI grant W911NF-14-1-0003 from ARO and by DOE grant de-sc0007870 and DOE Grant No. DE-SC0019030.  A.V. and E.K. were supported by a Simons Investigator award and by nsf-dmr 1411343.

\appendix
\begin{widetext}
\section{Derivation of the Hamiltonian}
\label{App:derivation}
Our setting is a generalization of the twisted bilayer problem considered in Refs.~\cite{Bistritzer2011,Santos2007}. We consider $n$ graphene layers where the $\ell$-th layer is rotated counter-clockwise by an angle $\theta_{\ell}$ then displaced by distance $\bd_{\ell}$ relative to a fixed reference and restrict ourselves to coupling only between next neighboring layers which can, in general, depend on the layer index $\ell$. Following Bistritzer and Macdonald \cite{Bistritzer2011}, the coupling between layers is given by the term
\beq
\label{tij}
T^{ij} = \int d\br d\br' c^\dagger_{i}(\br) t^{ij}(\br, \br') c_{j}(\br')\,,
\eeq
where $c_{\ell}(\br)$ represents the annihilation operator for an electron in the $\ell$-th layer at position $\br$. The electron operator in the $\ell$-th layer can be expanded in terms of the graphene orbitals living on a honeycomb lattice which is twisted by angle $\theta_{\ell}$ and displaced by $\bd_{\ell}$
\beq
\label{cr}
 c_{\ell}(\br) = \sum_{\bR,\alpha} \phi_{\bR,\alpha}(R_{-\theta_\ell} (\br - \bd_\ell)) f_{\ell,\bR,\alpha} = \sum_{\bR,\alpha} \phi(R_{-\theta_\ell} (\br - \bd_\ell) - \bR - \tau_\alpha) f_{\ell,\bR,\alpha}\,.
\eeq 
Here, $\phi_{\bR,\alpha}$ is the orbital centered at unit cell $\bR = m_1 \ba_1 + m_2 \ba_2$ in sublattice $\alpha$, $R_{\theta}$ is the rotation matrix $e^{-i \theta \sigma_y}$, and $\tau_\alpha$ is the real space position of the sublattice $\alpha$ given by $\tau_{A/B} = (\frac{\sqrt{3}}{2}, \pm \frac{1}{2}) a$. We now substitute in (\ref{tij}) and assume that the orbital $\phi_{\bR,\alpha}(\br)$ is strongly localized in space such that $\phi_{\bR,\alpha}(\br) \approx \delta(\br - \bR - \tau_\alpha)$ yielding
\beq
\label{tff}
T^{ij} = \sum_{\bR, \bR', \alpha, \alpha'} t^{ij}(R_{\theta_i} (\bR + \tau_\alpha) + \bd_i, R_{\theta_j} (\bR' + \tau_{\alpha'}) + \bd_j)  f^\dagger_{i,\bR,\alpha} f_{j,\bR',\alpha'}\,.
\eeq
Next, we assume $t^{ij}(\br, \br')$ only depends on $|\br - \br'|$ and introduce the Fourier transform
\beq
t^{ij}(\br, \br') = \sum_\bq e^{i \bq \cdot (\br - \br')} t^{ij}_\bq\,.
\eeq
In addition, we expand the operator $f_{i,\bR,\alpha}$ in terms of the annihilation operator for the Bloch states
\beq
f_{i,\bR,\alpha} = \sum_{\bp \in {\rm BZ}_i} e^{i \bp \cdot (\bR + \tau_\alpha)} \psi_{i,\bp,\alpha}\,.
\eeq
Substituting in (\ref{tff}) and using $t_{\bq} = t_{|\bq|}$ yields
\beq
\label{Tpsi}
T^{ij} = \sum_{\alpha, \alpha', \bG, \bG', \bp, \bp'} \delta_{R_{\theta_i} (\bG + \bp), R_{\theta_j} (\bG' + \bp')} t^{ij}(\bG + \bp) e^{i [\bG \cdot \tau_\alpha - \bG' \cdot \tau_{\alpha'} + R_{\theta_i} (\bG + \bp) \cdot (\bd_i - \bd_j)]} \psi^\dagger_{i,\bp,\alpha} \psi_{j,\bp',\alpha'}\,.
\eeq
Here, the momentum $\bp (\bp')$ is measured relative to the Brillouin zone center in the $i$-th ($j$-th layer).

Next, we consider the limit when both $\bp$ and $\bp'$ are close to the $K$ point and the rotation angles $\theta_i$ and $\theta_j$ are small. In this case, we can restrict the sum in (\ref{Tpsi}) to the largest terms which correspond to $\bp \approx K$ and $\bG = 0, \bG_2, \bG_3$ where $\bG_{2,3} = \left( - \frac{2\pi}{\sqrt{3} a}, \pm \frac{2\pi}{3 a} \right)$. The Bloch state annihilation operator $\psi_{i,\bp,\alpha}$ in the vicinity of $K$ can be expressed in terms of the annihilation operator for real space slowly varying orbitals at the K valley as
\beq
\label{psi}
\psi_{i,\bp,\alpha} = \int d\br e^{-i (\bp - K) \cdot \br} \tilde \chi_{i, K, \alpha}(\br)\,.
\eeq
Here $\tilde \chi_{i,K,\alpha}(\br)$ denotes the annihilation operator for an electron in valley K, sublattice $\alpha$ at point $\br$ measured relative to the coordinate system of the $i$-th layer. The same operator in the reference coordinate system is given by $\chi_{i,K,\alpha}(\br) = \tilde \chi_{i,K,\alpha}(R_{-\theta_i} (\br - \bd_i))$. Substituting in (\ref{psi}) gives
\beq
\psi_{i,\bp,\alpha} = \int d\br e^{-i (\bp - K) \cdot R_{-\theta_i} (\br - \bd_i)} \chi_{i, K, \alpha}(\br)\,.
\eeq
We now express the momenta close to the $K$ point in terms of the reference (unrotated Brillouin zone) $\bp = R_{-\theta_i} \bk + K$, $\bp' = R_{-\theta_j} \bk' + K$ to get
\begin{gather}
T^{ij} = \int d\br c^\dagger_{i,\alpha}(\br) [T^{ij}(\br)]_{\alpha,\alpha'} c_{j,\alpha'}(\br)\,, \nonumber \\ [T^{ij}(\br)]_{\alpha,\alpha'} = w^{ij} \!\!\! \sum_{\bG = 0, \bG_2, \bG_3} \!\!\! e^{-i(R_{\theta_i} - R_{\theta_j}) (\bG + K) \cdot \br} e^{i (\bG \cdot (\tau_\alpha - \tau_{\alpha'})+ R_{\theta_i} (\bG + K) \cdot \bd_i - R_{\theta_j} (\bG + K) \cdot \bd_j)}\,.
\label{Tij1}
\end{gather}
To show that this reduces to the Bistritzer Macdonald case, we perform the gauge transformation $c_i \rightarrow c_i e^{i R_{\theta_i} K \cdot \bd_i}$ which leads to
\beq
T^{ij} \rightarrow T^{ij} e^{-i (R_{\theta_i} K \cdot \bd_i - R_{\theta_j} K \cdot \bd_j)} = w^{ij} \!\!\! \sum_{\bG = 0, \bG_2, \bG_3} \!\!\! e^{-i(R_{\theta_i} - R_{\theta_j}) (\bG + K) \cdot \br} e^{i (\bG \cdot (\tau_\alpha - \tau_{\alpha'})+ R_{\theta_i} \bG \cdot \bd_i - R_{\theta_j} \bG \cdot \bd_j)}
\eeq
which reduces to the expression of Bistritzer and Macdonald for $R_{\theta_i} = 1$, $\bd_i = 0$. In the following, we will prefer to use the form (\ref{Tij1}) because the way displacement enters in the Hamiltonian is more transparent as we see below. We now assume that the value of $w^{ij}$ is different between the diagonal and off-diagonal terms to take into account the lattice relaxation effects \cite{Carr2019} leading to
\begin{gather}
T^{ij}(\br) = \left(\begin{array}{cc} 
w^{ij}_{\rm AA} U_0^{ij}(\br) & w^{ij}_{\rm AB} U_1^{ij}(\br) \\ 
w^{ij}_{\rm AB} U_1^{ij*}(-\br) & w^{ij}_{\rm AA} U_0^{ij}(\br) 
\end{array} \right)\,, \\
U^{ij}_m(\br) = \sum_{n=1}^3 e^{i m (n-1) \phi} e^{-i \bq_n^{ij} \cdot (\br - \bD_{ij})} , \quad m=0,1, \quad \bq^{ij}_1 = 2 k_D \sin (\theta_{ji}/2) R_{\phi_{ij}}(0, -1), \quad \bq^{ij}_{2,3} = R_{\pm \phi} \bq^{ij} \\
\phi = 2\pi/3, \quad \theta_{ji} = \theta_j - \theta_i, \qquad \phi_{ij} = (\theta_i + \theta_j)/2, \qquad \bD_{ij} = \frac{\bd_i + \bd_j}{2} + i\cot (\theta_{ji}/2) \sigma_y \frac{\bd_i - \bd_j}{2}
\end{gather}
and $  k_D = \frac{4\pi}{3 \sqrt{3} a}$. The full Hamiltonian can then be written as
\beq
H = \sum_{\ell=1}^n c_{\ell,\br}^\dagger \cdot (-i v_F  \bsigma_{\theta_{\ell}}  \bnabla) c_{\ell,\br} + \sum_{\ell=1}^{n-1} c_{\ell,\br}^\dagger T^{\ell,\ell+1}(\br) c_{\ell+1,\br} + \text{h.c.}\,,
\eeq
where $\bsigma_{\theta_{\ell}} = e^{\frac{i}{2} \theta_{\ell} \sigma_z} \bsigma e^{-\frac{i}{2} \theta_{\ell} \sigma_z}$. The (first quantized) Hamiltonian can be written explicitly as
\begin{align}
\H = \left(  \begin{array}{ccccc}
  -iv_{F}  \bm{\sigma}_{\theta_{1}}  \bm{\nabla} & T^{12}(\br) & 0 & \dots & 0 \\ 
    T^{12\dag}(\br)  &  -iv_{F} \bm{\sigma}_{\theta_{2}}  \bm{\nabla} & T^{23}(\br) & \dots & 0 \\ 
    0 & T^{23\dag}(\br) & -iv_{F}\bm{\sigma}_{\theta_{3}}  \bm{\nabla} & \dots & 0 \\ 
    \dots & \dots  & \dots & \ddots & T^{n-1,n}(\br) \\ 
    0 & 0 & 0 & T^{n-1,n\dag}(\br)  &   -iv_{F} \bm{\sigma}_{\theta_n}  \bm{\nabla} \\ 
  \end{array}\right)\,.
\end{align}

\end{widetext}

\bibliography{refs}

\end{document}